\documentclass[aps,showpacs,nofootinbib, 11pt]{revtex4-1}
\usepackage{latexsym}
\usepackage{amsmath,amsfonts, MnSymbol}
\usepackage{amsbsy}
\usepackage{mathrsfs}
\usepackage{verbatim}
\usepackage{graphicx,calc,epsfig}
\def\ut#1{\rlap{\lower1ex\hbox{$\sim$}}#1{}}
\def\sdpb#1{\rlap{\lower1.5ex\hbox{$\Leftarrow$}}{#1}}

\newcommand{\C}{\mathbb{C}}

\newcommand{\Z}{\mathbb{Z}}
\newcommand{\be}{\nopagebreak[3]\begin{equation}}
\newcommand{\ee}{\end{equation}}
\newcommand{\bee}{\nopagebreak[3]\begin{equation*}}
\newcommand{\eee}{\end{equation*}}
\newcommand{\ba}{\nopagebreak[3]\begin{eqnarray}}
\newcommand{\ea}{\end{eqnarray}}
\newcommand{\baa}{\nopagebreak[3]\begin{eqnarray*}}
\newcommand{\eaa}{\end{eqnarray*}}
\newcommand{\la}{\label}
\DeclareFontFamily{U}{rsfs}{}         
\DeclareFontShape{U}{rsfs}{m}{n}{<5> rsfs5 <6><7> rsfs7          %
  <8><9><10><10.95><12><14.4><17.28><20.74><24.88> rsfs10}{}     %
\DeclareMathAlphabet{\mathfs}{U}{rsfs}{m}{n}                     %
\newcommand{\mfs}[1]{\mathfs {#1}}                               %
\newcommand{\va}{\scriptscriptstyle}

\newcommand{\van}{\scriptstyle}

\newcommand{\sH}{{\mfs H}}

\newcommand{\sI}{{\mfs I}}

\def\i{i}

\begin{document}

\title{Dynamical evaporation of quantum horizons}

\date{\today}

\author{Daniele Pranzetti$^1$}

\thanks{pranzetti@aei.mpg.de}

\affiliation{$^1$Max Planck Institute for Gravitational Physics (AEI), 
Am M\"uhlenberg 1, D-14476 Golm, Germany.}


\begin{abstract}

We describe the black hole evaporation process driven by the dynamical evolution of the quantum
gravitational degrees of freedom resident at the horizon, as identified by the loop quantum gravity
kinematics. Using a parallel with the Brownian motion, we interpret the first law of quantum
dynamical horizon in terms of a fluctuation-dissipation relation. In this way, the horizon evolution is described in terms of relaxation to an equilibrium
state balanced by the excitation of Planck scale constituents of the horizon. This discrete quantum hair structure associated to the horizon geometry produces a deviation from thermality in the radiation spectrum.
We investigate the final
stage of the evaporation process and show how the dynamics leads to the formation of a massive
remnant, which can eventually decay. Implications for the information paradox are discussed.

\end{abstract}


\maketitle


\section{Introduction} \la{Introduction}

In the late sixties, early seventies a fascinating parallel between black holes physics and thermodynamics started to be delineated, which culminated in the derivation of four laws analog to those of a thermodynamical system \cite{Bekenstein, Bardeen}. However, this analogy was not originally taken too seriously by many since a black hole has classically zero temperature. A simple dimensional analysis shows how, in order to talk about black hole temperature, one needs to refer to the quantum theory. This motivated Hawking's seminal study of a quantum scalar field on a Schwarzschild background, which led to the discovery of black holes evaporation \cite{Hawking}. However, as soon realized by Hawking himself, this result, together with the singularity theorems \cite{Hawking-Ellis, Wald}, leads to a violation of unitary evolution \cite{Hawking2}. All these elements represent a quite strong evidence of the need of a quantum treatment of the gravitational field in order to fully understand black holes physics. 

In this regard, the main points a quantum theory of gravity should address and explain are: the microscopic origin of the entropy degrees of freedom and the dynamics of the evaporation process. While the former has received a lot of attention in the last twenty years, with several proposals within different approaches, more or less successful in reproducing the semi-classical result of Bekenstein and Hawking, the latter has been object of much less investigation.


In this letter, we want to concentrate on the description of the dynamical evaporation of quantum horizons, as recently introduced in the context of the Loop Quantum Gravity (LQG) approach \cite{H-radiation}, and analyze its implications for the information paradox. 

The main idea at the core of our analysis is the notion of quantum horizon as a gas of particles. Such a notion can be traced back to an old proposal by Bekenstein \cite{Bekenstein2} and it has been exploited within the years by different authors (see, e.g., \cite{BM, Krasnov, H-radiation}). We believe that the main motivations in support of such an analogy come from the points of view championed, among others, by Sorkin \cite{Sorkin}, Smolin \cite{Smolin2}, Jacobson and Rovelli \cite{Jacobson, Rovelli, JMR} on the nature of black hole entropy. In particular, these authors argue that the entropy in the first law is the logarythm of the number of states of the black hole that can affect the {\it exterior} and such degrees of freedom have to reside on the horizon; moreover, the finiteness of the entropy has to be related to the deep, discrete structure of space-time. 

In the LQG picture, the information on the horizon accounting for the entropy is stored in the spin network links piercing the horizon \cite{ABCK, SU(2), SU(2)2, Review}, which encode the quantum fluctuations of the geometry: the `charge' associated to the topological defects on the boundary correspond to a {\it quantum hair} for the black hole. 
Here, we want to exploit further this picture of a quantum horizon as a discrete system whose constituents represent the `atoms' of space. 

In analogy to Einstein's treatment of Brownian motion, 
we show that the first law of quantum dynamical horizons \cite{DH1, DH2}, based on the loop quantization of the Hamiltonian constraint, can be interpreted as a fluctuation-dissipation theorem for the horizon degrees of freedom. In this way, the  radiation spectrum produced by the action of the Hamiltonian operator near the horizon could play a role, analog to that of Einstein's relation for the atomic theory of matter, in proving the atomic structure of quantum space, as described by the kinematical Hilbert space of LQG. This is the first main result of the paper and will be presented in
Section \ref{Sciama}. Here we also explore the relation between the microscopic action of the Hamiltonian operator and the emergent macroscopic description of the horizon dynamics.

In Section \ref{paradox} we investigate the implications of the existence of this quantum hair at the horizon (associated to the quantum geometry d.o.f.) and of possible non-local effects in the quantum gravitational regime of the collapse to describe the last stage of evaporation process within the LQG framework. By implementing locally the dynamics encoded in the Hamiltonian operator till the horizon reaches a Planck scale size, we show that the horizon area operator result to be bounded from below, with the minimum eigenvalue allowed by the dynamics being $8\pi \beta \ell_p^2 \sqrt{2}$. The analysis shows that
{\it conservative} scenarios can be realized, leading to either the formation of a massive remnant or the dissolution of the horizon. 
This is the second main result obtained here and it provides support to a singularity-free evolution from the full theory and a possible solution to the information paradox.  

Conclusions are presented in Section \ref{Conclusions}.

\section{Fluctuation-dissipation theorem}\la{Sciama}

Fluctuation-dissipation theorems represent very powerful tools to study the fluctuations of systems described by statistical mechanics. They express the existence of a relation between the spontaneous fluctuations and the response to external fields of physical observables. Fluctuation-dissipation theorems are based on Onsager's principle in the theory of dissipative processes, stating that a linear system behaves on average in the same way in a given configuration whether it reached that
configuration by a spontaneous fluctuation or by an externally induced perturbation. Motivated by a suggestion of Candelas and Sciama \cite{Sciama} to understand Hawking radiation in these terms, the main result of this section is to show how, within the dynamical horizons framework \cite{DH1, DH2} and a local statistical mechanical perspective \cite{APE}, a physical process version of the first law can be understood as a fluctuation-dissipation theorem and used to describe the evaporation process. Such an area balance law can be written in terms of the Hamiltonian constraint of gravity. Then, the LQG quantization prescription for the Hamiltonian operator implemented locally near the horizon is used to relate the microscopic description of the geometry fluctuations to the radiation spectrum and the dissipative nature of the evaporation process. In this way, black hole radiance is described from a completely new perspective, namely from a local quantum gravity point of view. At the same time, this new level of analysis provides a new, deeper statistical mechanical understanding of black holes as thermodynamical systems.


In order to present these ideas in a clearer and more pedagogical way, we first review in section \ref{BM} a specific example of fluctuation-dissipation theorem successfully applied to connect the macroscopic and microscopic levels of description of a physical system, namely the Brownian motion. The new result is presented in section \ref{diss}.
We conclude in section \ref{Ilaw} with some speculative observations relating a modified first law for IH recently proposed in \cite{AP}, a generalized Clausius law 
taking into account non-equilibrium dissipative processes and a coarse grained description of the Hamiltonian operator action emerging from the fluctuation-dissipation theorem interpretation.

\subsection{Brownian motion}\la{BM}

The first example of fluctuation-dissipation theorems was provided by Einstein's work on Brownian motion, which culminated with the famous `Einstein relation'
\be\la{Einstein}
D=\mu k T,
\ee
expressing the diffusion coefficient $D$ of a Brownian particle in terms of its mobility $\mu$, through the temperature $T$ of the fluid and Boltzmann's constant $k$. The experimental verification of Einstein relation allowed the determination of the Avogadro number (a microscopic quantity) from accessible macroscopic quantities, thus providing conclusive evidence of the existence of atoms.

In order to understand how (\ref{Einstein}) encodes a relation between fluctuations and the response to external perturbations, let us quickly go through its derivation. Einstein's analysis can be divided into two parts. The first part of his argument consisted of considering the collective motion of Brownian particles and showing that the particle density $n(x,t)$ satisfies the diffusion equation
\be
\frac{\partial n}{\partial t}=D\frac{\partial^2 n}{\partial x^2},
\ee
from which the mean square displacement grows in proportion to time according to
\be\la{BM1}
\langle x(t)^2\rangle =2Dt.
\ee
The second part of Einstein's theory involves a dynamic equilibrium established between opposing forces and is what the fluctuation-dissipation theorem arises from. Following Langevin's approach, one can write down a stochastic differential equation taking into account the effect of molecular collisions by means of an average force, given by the fluid friction, and of a random fluctuating term, namely
\be\la{sto}
m\frac{d v}{dt}=-m\gamma v + R(t),
\ee
where, assuming Stokes law for the frictional force $f_v=-m\gamma v$, $m\gamma=6\pi a \eta$, with $a$ the particle radius and $\eta$ the fluid viscosity. The component $R(t)$ of the force resulting from the action of the molecules of the fluid on the Brownian particle is a random fluctuating force, independent of the particle motion. By means of Fourier analysis applied to the random force $R(t)$ and the velocity $v(t)$ of the Brownian particle, the stochastic differential equation \eqref{sto} can be written as
\be\la{sto2}
v(\omega)=\frac{1}{\i\omega+\gamma}\frac{R(\omega)}{m},
\ee
where $v(\omega), R(\omega)$  are the Fourier modes. By means of the Wiener-Khintchine theorem that associate the correlation function of a given process $z(t)$ to its intensity spectrum via
\be\la{WK}
I(\omega)=\frac{1}{2\pi}\int_{-\infty}^\infty \langle z(0) z(t)\rangle e^{-\i\omega t}\,,
\ee
eq. \eqref{sto2} gives
\be\la{sto3}
I_u(\omega)=\frac{1}{\omega^2+\gamma^2}\frac{I_R(\omega)}{m^2}\,.
\ee
Hence, knowing the power spectrum $I_R(\omega)$, eq. \eqref{sto3} converts it into $I_u(\omega)$, allowing to solve the initial Langevin equation \eqref{sto} to the same extent.

Now, let us assume for simplicity that the power spectrum $I_R$ of the random force $R(t)$ is independent of frequency and demand  the equipartition law $m\langle v^2\rangle=kT$ to hold also for the colloidal particle, as expected if this is kept for a sufficiently long time in the fluid. Then it follows from \eqref{sto3} with \eqref{WK} that 
 the response function (mobility) of the system $\mu=(m\gamma)^{-1}$ can be associated to the correlation function of the stochastic process $v(t)$ through the relation
\be\la{BM2}
\mu=\frac{1}{k T}\int_0^\infty \langle v(0) v(t)\rangle dt.
\ee

As a last step, we write the mean square average of the displacement of the Brownian particle in a time interval $(0,t)$ as
$\langle x(t)^2\rangle=\int_0^t dt_1\int_0^t dt_2 \langle v(t_1) v(t_2)\rangle$
and transform this into
\be
\lim_{t\rightarrow\infty} \frac{\langle x(t)^2\rangle}{2t} =\int_0^\infty \langle v(0) v(t)\rangle dt\,.
\ee
Thus, combining the result \eqref{BM1} of the first part of the analysis with the fluctuation-dissipation theorem \eqref{BM2}---which links the macroscopic (validity of the Stokes law) and microscopic (assumption that the Brownian particle is in statistical equilibrium with the molecules in the liquid) levels of description---yields immediately the Einstein relation \eqref{Einstein}.

\subsection{Horizon dissipative processes}\la{diss}

A couple of years after Hawking's derivation of black holes radiance \cite{Hawking}, Candelas and Sciama \cite{Sciama} applied the concepts behind Onsager's principle to study the thermodynamics of dissipative processes associated to black holes physics. The main idea of Candelas and Sciama was to derive Hawking radiation by means of a fluctuation-dissipation theorem relating the black hole area dissipation rate to the fluctuations of a quantum shear operator associated to gravitational degrees of freedom on the horizon. In order to do so, in analogy to the Brownian motion example, they studied the effect of a gravitational perturbation encoded in a non-vanishing shear $\sigma$ by analyzing the spontaneous vacuum fluctuations of the shear itself. 

Starting from the Hawking-Hartle relation \cite{Hartle} for an horizon area increase in presence of a purely gravitational stationary perturbation at lowest order---expressing, for example, the rate of slowing down of a black hole by a non-axisymmetric gravitational field produced by distant masses---, Candelas and Sciama interpret $\sigma$ as a quantum operator and write down the following fluctuation-dissipation relation
\be\la{HH}
\frac{d A}{dt}=\frac{2}{\kappa}\int \langle \sigma^2 \rangle dA,
\ee
where $t$ is a suitably defined time variable on the horizon and $\kappa$ the surface gravity. By properly choosing the vacuum state, they compute the r.h.s. of \eqref{HH} and shows that the shear fluctuations have the stochastic properties of black-body radiation at temperature $\kappa/2\pi$. Therefore, as the perturbation is dissipated away and a stationary state is approached, the horizon would emit gravitational radiation matching Hawking's result. 

Such a simple and elegant derivation of black holes radiance shows the power and usefulness of flcutuation-dissipation theorems in studying non--equilibrium statistical mechanics. We now want to interpret our analysis in \cite{H-radiation} of the radiation process along those lines and make the idea behind the implementation of the near-horizon dynamics introduced there more precise. By doing so, we will argue how, in analogy to Einstein's work on the Brownian motion, black holes evaporation could provide a proof for the atomic structure of quantum space.

In \cite{H-radiation}, we applied the dynamical horizons formalism developed in \cite{DH1, DH2} to study the transition between two equilibrium configurations of the horizon\footnote{See \cite{Amit} for an application of the dynamical horizons formalism in the context of Hawking radiation.}. More precisely, together with the local statistical mechanical framework introduced in \cite{AP, APE}, a physical process version of the first law derived in \cite{DH2} has been used to implement the bulk dynamics near the horizon, as described by the LQG approach, and evolve the horizon quantum geometry.

Such a first law was derived from an {\it area balance law} relating the change in the area of the dynamical horizon to the flux of matter and gravitational energy. In the vacuum, for a non-rotating horizon the canonical form of Einstein equation gives \cite{DH2}
\be\la{balance1}
\frac{1}{16 \pi G}\int_{\va \Delta \sH}\!\! N_r H d^3V=\int_{r_1}^{r_2} \frac{dr}{2G}- \frac{1}{8 \pi G}\int_{\va \Delta \sH}\!\! N_r \sigma^2 d^3V=0,
\ee
where $\Delta \sH$ is the portion of dynamical horizon bounded by the 2-sphere leaves $S_1, S_2$ of radius $r_1, r_2$, $N_r$ a lapse function, $H$ is the scalar Hamiltonian constraint and $\sigma$ is the shear of a null vector field $\ell^a$ normal to the leaves $S$ foliating the horizon $\sH$. From eq. \eqref{balance} then one obtains the analog of \eqref{HH}, namely
\be\la{balance}
\frac{ \kappa_r}{8\pi G} \frac{dA}{dr}= \frac{1}{8 \pi G}\int_S\!\!  \sigma^2 d^2V,
\ee
where $A=4\pi r^2$, $\kappa_r$ is the surface gravity associated with the vector field $\xi^a_r=N_r\ell^a$ and the lapse has been chosen such that $d^3V=N_r^{-1}drd^2V$. The dynamical version of the first law \eqref{balance} relates the infinitesimal change of the horizon area in `time' (played by the radial coordinate $r$ along $\Delta \sH$) to the flux of gravitational energy associated with $\xi^a_r$. One can then use the freedom to reparametrize the time variable $r$ with an arbitrary function $f(r)$---encoding the freedom to rescale the vector field $\xi_r$---to identify the l.h.s of \eqref{balance} with the local notion of horizon energy introduced in \cite{APE}. Namely, by choosing the function $f(r)$ as the proper distance $\rho$ of a preferred family of static observers hovering closely outside the horizon\footnote{Notice that this choice corresponds to the lapse=1 gauge, as can be easily verified from the relation for the rescaling of the lapse function $N_\rho=\frac{d\rho}{dr} N_r$.}, one gets
\be\la{Edot}
\dot E\equiv \frac{dE}{d\rho}=\frac{1}{8 \pi G}\int_S\!\!  \sigma^2 d^2V,
\ee
where $E= \kappa_{\va \rho} A/8\pi G$ and $\kappa_{\va \rho}=(d\rho/dr)  \kappa_r=1/\rho+o(\rho)$ represents the local surface gravity measured by the stationary observer. Considering the local perspective we are assuming here, the proper distance observable emerges as a natural choice, corresponding to a Rindler form of the near-horizon metric. In fact, it is the only choice such that, once a stationary equilibrium configuration is reached again, the energy associated to the corresponding Killing vector field matches the physical notion of energy derived in \cite{APE}.

Relation \eqref{Edot} can now be used in the quantum theory to study the spectrum of the evaporation process, in analogy to \cite{Sciama}. The dynamical phase can be studied as a perturbation between two equilibrium states represented by the IH configurations. Classically, isolated horizons are defined as null internal boundaries of space-time whose congruence of null generator vector fields has vanishing expansion (see \cite{IH} for a detailed definition) plus some energy condition. From this definition one can show that on each 2-sphere foliating the horizon the following boundary conditions hold
\be\la{eom}
{F}(A) = -\frac{\pi (1-\beta^2)}{a_{\va
H}}\,{\Sigma},
\ee
where $A$ is the Ashtekar-Barbero connection, $\Sigma$ the 2-form dual of the densitized triad conjugate to $A$, $a_{\va
H}$ the horizon area and $\beta$ the Barbero-Immirzi parameter. At the quantum level then the kinematical Hilbert space can be split in a bulk and a surface part.
The bulk space geometry is described by the polymer-like excitations of the gravitational field encoded in the spin networks states, which span the kinematical Hilbert space of LQG. Some edges of those states can now pierce through the horizon surface, providing local quantum d.o.f. accounting for the horizon entropy. More precisely, for a fixed graph $\gamma$ in the bulk $M$ with end points on the isolated horizon $IH$, denoted $\gamma\cap IH$, the quantum operator associated with $\Sigma$ in (\ref{eom}) is
\begin{equation}
\label{gammasigma} \epsilon^{ab}\hat{\Sigma}^i_{ab}(x) = 16 \pi G
\beta \sum_{p \in \gamma\cap IH} \delta(x,x_p) \hat{J}^i(p)\,
\end{equation}
where $[\hat{J}^i(p),\hat{J}^j(p)]=\epsilon^{ij}_{\ \ k} \hat{J}^k(p)$ at each $p\in\gamma\cap IH$.
let us denote $|\{j_p,m_p\}_{\va 1}^{\va n}; {\van \cdots} \rangle$ the boundary state, where $j_p$ and $m_p$ are the spins and magnetic numbers labeling the $n$ edges puncturing the horizon
at points $x_p$ (other labels are left implicit). The horizon area operator $\hat a_{\va H}$ is diagonal on this state namely 
\be
\hat a_{\va H}|\{j_p,m_p\}_{\va 1}^{\va
n}; {\van \cdots} \rangle=8\pi\beta \ell_p^2 \,
\sum_{p=1}^{n}\sqrt{j_p(j_p+1)} |\{j_p,m_p\}_{\va 1}^{\va
n}; {\van \cdots} \rangle.
\ee
The boundary theory then is quantized as a Chern-Simons theory in presence of particles. In fact, by defining 
$ k\equiv a_{\va H}/(4\pi \ell_p^2\beta (1-\beta^2))$, the quantum boundary condition \eqref{eom} can be rewritten as
\be\la{CS}
\hat F(A) = \frac{4\pi}{k} \sum_{p \in \gamma\cap IH} \delta(x,x_p) \hat{J}^i(p)
\ee
where we have identified the $\hat{J}^i(p)$ LQG operators with the with the Chern-Simons particles spin oprators.
From the eom \eqref{CS} we see that the curvature of the Chern-Simons connection vanishes everywhere on $IH$ except at the position of the defects where we find conical singularities of strength proportional to the defects' momenta \cite{SU(2)2}. Moreover, there is an important global constraint that follows from \eqref{CS} implying that the Chern-Simons boundary Hilbert space be isomorphic to the $SU(2)$ singlet space between all the punctures, once the large horizon area limit is taken. In this way, one recovers the $SU(2)$ intertwiner model of \cite{SU(2)2} (see FIG. \ref{IHfig} below).

\begin{figure}[ht]
\centering
\includegraphics[scale=0.4]{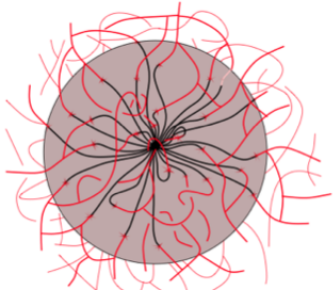}
\caption{Boundary Hilbert space represented by a single $SU(2)$ intertwiner.}
\label{IHfig}
\end{figure}

The IH boundary conditions imply that lapse must be zero at the horizon so that the Hamiltonian constraint is only imposed in the bulk. Now we want to `unfrozen' the bulk dynamics near the horizon by interpolating two IH configuration with a dynamical horizon phase $\Delta \sH$, as studied in \cite{DH2}. As shown above, the Hamiltonian constraint $\int_{\va \Delta \sH}\! N_r H d^3V=0$ takes the form \eqref{Edot} in the gauge $f=\rho$, corresponding to a local stationary observer point of view.
In the quantum theory then the (deparametrized) version of eq. \eqref{balance1} can formally be written as
\be\la{Scalar}
\hat H =  \hat p_\rho + \hat H_0=0,
\ee
where $ \hat p_\rho=\Delta \hat E$, given by the (variation of the) area Hamiltonian, plays the role of the momentum conjugate to the time variable $\rho$ and corresponds to the energy of the emitted quantum of radiation; $ \hat H_0=\hat{ \sigma}^2/8 \pi G$ is the Hamiltonian related to a shear operator driving the area variation, hence evolving the boundary states\footnote{In this way, the dynamical horizons framework allows us to adopt an approach to study the evaporation process close in spirit to the one of \cite{Massar}. 
}. Notice that in this dynamical context only $ \hat p_\rho=\Delta \hat E$ is a Dirac observable
and the imposition of the Hamiltonian constraint \eqref{Scalar}, i.e. the implementation of dynamics of the theory, corresponds to a relation between {\em partial observable} $\rho$ and $\hat E$ \cite{Observables}. More precisely,
going to the Heisenberg picture, the dynamics encoded by \eqref{Scalar} can be used to study the radiation spectrum induced by the dissipation of the horizon energy by means of the matrix elements of the Hamiltonian constraint operator, since (the matrix elements of) the dissipation rate of the horizon energy observable can be written as
\be\la{quantumE}
\langle f| \dot{ \hat E}|i \rangle=  \langle f| [ \hat E, \hat H ] |i\rangle=
 \Delta  E \langle f| \hat H |i\rangle,
\ee
where   $|i\rangle\equiv|\{j^{\va i}_p,m^{\va i}_p\}_{\va 1}^{\va n^{\va i}}; {\van \cdots} \rangle , |f\rangle\equiv|\{j^{\va f}_p,m^{\va f}_p\}_{\va 1}^{\va n^{\va f}}; {\van \cdots} \rangle $ are the initial and final sets of punctures data, defining the eigenstates of the area Hamiltonian before and after the action of $\hat H$.

Eq. \eqref{quantumE} allows us to relate the horizon energy dissipation rate to the spectrum 
of the full Hamiltonian operator. Let us explain more in detail the meaning of this proposal for the implementation of the near-horizon dynamics.

At the quantum level, the scalar constraint acts locally at the vertices of the spin network states and changes the spin associated to the edges attached to the given vertex, hence inducing fluctuations of the quantum geometry. Using the kinematical picture defined above, if we consider
Thiemann's proposal \cite{QSD} and concentrate only on the Euclidean part for simplicity, the action of the Hamiltonian constraint on a 3-valent node having two edges piercing the horizon can be graphically represented as
\be
\hat H\triangleright~\begin{array}{c}   \includegraphics[width=2.7cm,angle=360]{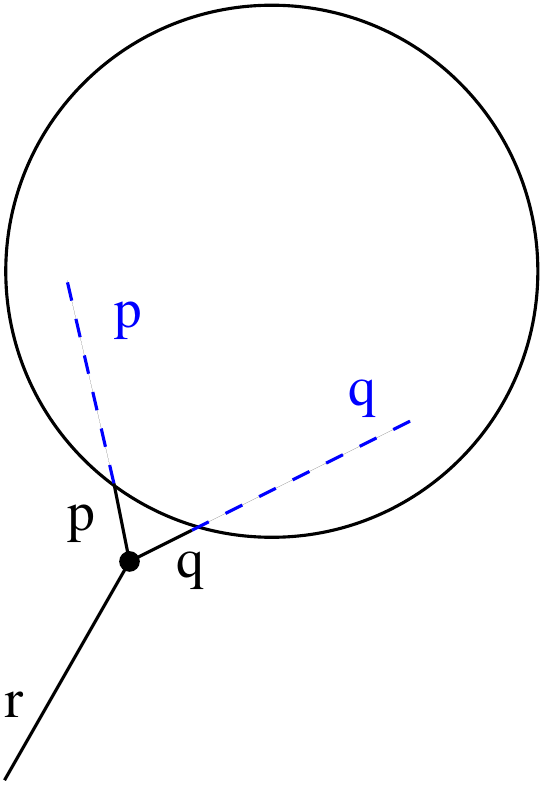}\end{array}~~~~~~\Rightarrow~~~~~\begin{array}{c}  \includegraphics[width=3cm,angle=360]{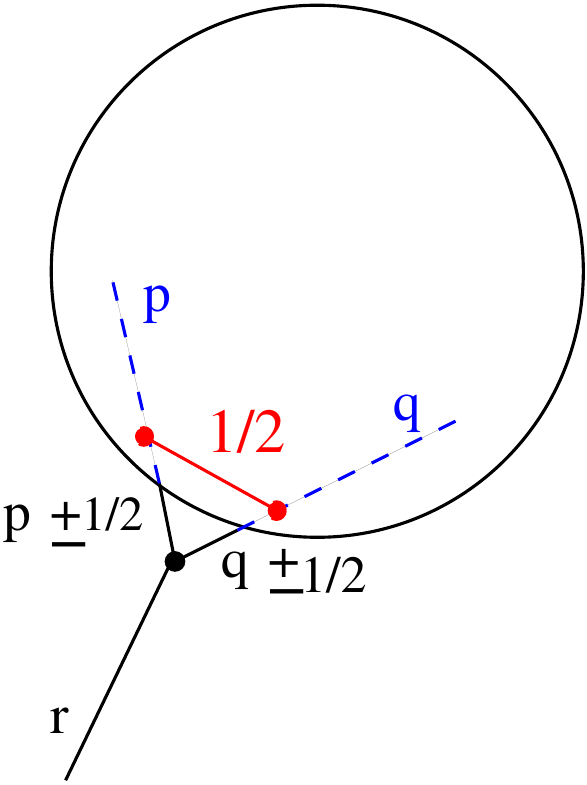}\end{array}\!,\la{H}
\ee
where the holonomies entering the regularization of $\hat H$ are taken in the fundamental representation.
The action of the Hamiltonian operator $\hat H$ near the horizon makes the initial horizon area operator eigenstate $|i\rangle$ on the left of \eqref{H} (the other punctures forming the boundary Hilbert space and not affected by the action of $\hat H$ are not shown in the picture) jump to a different one $|f\rangle$ on the right, 
generating in this way a change in the local notion of horizon energy and hence producing radiation\footnote{The action of $\hat H$ on nodes far from the horizon gives a vanishing commutator with the observable $\hat E$, leaving the horizon state unchanged.}. The spectrum of such an emission process results in a discrete set of lines depending on the matrix elements of the Hamiltonian operator. The imposition of the scalar constraint \eqref{Scalar} on a (space-like) portion of dynamical horizon connecting the two stationary configurations encodes a relation between the horizon energy dissipation and the metric fluctuations induced by a near-horizon geometrical operator during the transition phase between the two consecutive equilibrium states. In the LQG description, the action of the full Hamiltonian operator on a vertex near the horizon \eqref{H} can be used to `evolve' the isolated horizon, through a dynamical phase, from one quantum configuration to another.
This is how relation \eqref{quantumE} should be understood.

We can now see the fluctuation-dissipation theorem interpretation of the first law of dynamical horizons associated to the relation \eqref{Edot}. The Hamiltonian constraint encodes the same characteristics of the force driving the Brownian motion: it's action induce both a frictional and a fluctuating effect on the horizon quantum geometry, with the `viscous' dissipation rate of the horizon area (energy) related to its matrix elements. The dissipative effect of the quantum fluctuation of geometry is related to the form of the area operator spectrum in LQG; in fact, due to the decreasing gap between higher eigenvalues, the overall balance of the geometry fluctuations results in a rate between energy emission and absorption bigger than one. In this way, the quantum theory provides a microscopic explanation for the dissipative nature of the evaporation process.

In this picture then, an eventual observation of black holes radiance could provide a macroscopic window on the microscopic world, playing a role analog to that of Einstein's relation in proving the existence of atoms. In fact, from such an observation one could derive the number of punctures of a given spin defining the quantum horizon geometry by means of the spectrum intensity relation analog of the Fermi golden rule
\be\la{spectrum}
I_{\va pq}=\bar s_p \bar s_q |\langle \hat H\rangle|^2 \Delta E^3_{\va pq}\,,
\ee
where transition probabilities are computed using \eqref{quantumE}.
In the previous equation $\bar s_j$ is the expectation value for the occupation number of punctures with a given spin $j$, $\langle \hat H\rangle$ are the LQG Hamiltonian operator matrix elements corresponding to transition amplitudes involving two punctures with spins $p$ and $q$ piercing the horizon  and jumping to a different energy level, with
$\Delta E_{\va pq}$ being the energy emission in this single process   (the plot of the relevant lines of the emission spectrum \eqref{spectrum} can be found in \cite{H-radiation}, where we used the matrix elements of $\hat H$ computed in \cite{Matrix}). The formula \eqref{spectrum} represents the spectrum of Hawking radiation encoding quantum gravity effects.

\subsection{Modified first law}\la{Ilaw}

The interpretation at the quantum level of the first law (\ref{balance}) as a fluctuation-dissipation theorem reveals interesting analogies  with
 the derivation of the Einstein equation from a non-equilibrium thermodynamical treatment of the gravitational d.o.f. performed in \cite{Shear, Chirco} and, at the same time, sheds light on it. Here it was shown that, in the case of non-vanishing shear at the horizon, the thermodynamical argument can still be run as long as non-equilibrium consideration are applied. More precisely, the shear contribution in the Raychaudhuri equation leads to an entropy balance relation in which purely geometrical d.o.f. are encoded in entropy production terms associated to gravitational energy fluxes. The presence of this internal entropy term leads to a generalized Clausius relation of the form \cite{Chirco}
\be\la{Clau}
dS=dS_{ex}+dS_{in}=\frac{\delta Q}{T}+\delta N,
\ee
where the $\delta N$ term is related to the exictation of internal/purely gravitational d.o.f. whose macroscopic effect is encoded in the horizon shear viscosity. 
The association of this internal entropy contribution to some sort of viscous work on the microscopic d.o.f. of the system goes along with the description of the Hamiltonian operator action emerging from the fluctuation-dissipation theorem interpretation provided above. In fact, in the quantum microscopic theory, this non-equilibrium entropy contribution is related to the area dissipation effect associated to the elimination of a puncture from the horizon, providing a natural relation with the modified first law for IH
\be\la{first}
dE=TdS+\mu dN
\ee
 proposed in \cite{AP}, where the extra term on the r.h.s. is added in order to take into account, in the quantum geometry description of the horizon, the presence of the quantum hair associated to the number of punctures, as mentioned in the Introduction (more details on this follow below). The modification \eqref{first} can be understood as a non-equilibrium, dissipative contribution introduced by the excitation of the quantum gravitational d.o.f.: it contains a quantity playing the role of a chemical potential conjugate to the number of punctures and can therefore be related to the evaporation process described in \cite{H-radiation}. 
 
 Therefore, the analogy between the thermodynamical approach in the derivation of Einstein equation in presence on non-equilibrium processes \cite{Shear, Chirco} and the dynamical evaporation of quantum horizons described in \cite{H-radiation} provides important insights into the proposal \eqref{first}. In fact,
it suggests an interpretation of this extra term as an internal entropy production term, i.e. an entropy associated to the dynamics of the quantum gravitational d.o.f. and encoded in the correlations between the radiation and the internal state of the horizon. More precisely, as described more in detail below, the action of $\hat H$ induces the disappearance of a puncture inside the horizon, representing a bit of information no more available to the external observer. In this way, the radiation process creates correlations between fluctuations of quantum geometry just outside and inside the horizon, with emission of quanta associated to the creation of new links in the hole bulk,
 hence providing an entanglement entropy contribution\footnote{The derivation of an analog of the generalized second law related to such mechanism, for instance along the lines of \cite{Sorkin, Sorkin2}, would be desirable to make this picture more physically robust and is left for future investigation.}. Such an interpretation could be made more precise by 
 (and at the same time could provide important insights into) the relation between the Boltzmann \cite{Rovelli, ABCK} and the von Neumann \cite{LT, Bianchi2} derivations of black hole entropy in LQG recently found in \cite{Temp-IH}. 
There the two pictures have been shown to be two equivalent and complementary descriptions of the horizon degrees of freedom.

 Furthemore, the analogy strengthens further the macroscopic, coarse grained interpretation of the Hamiltonian operator action on a quantum horizon as a non-equilibrium dissipative process via gravitational/microscopic d.o.f.. Such an hydrodynamics intuition also suggests a parallel with the {\it membrane paradigm} \cite{Maggiore} and the {\it stretched horizon} picture \cite{Stret}, providing a precise characterizations of the notion of ``atoms'' used in that context---the analysis that follows presents some analogies with the scenario depicted in \cite{Stret}, even though the framework we work in is quite different.


\section{Information paradox}\la{paradox}

Right after his discovery of black holes radiance, Hawking himself realized \cite{Hawking2} that such a phenomenon would lead to a breakdown of unitary evolution of black holes. The problem, usually referred to as the {\it information paradox} (see \cite{Preskill, Mathur} for some reviews), is two-fold. A first loss of information concerns the matter degrees of freedom that collapsed and formed the black hole. This is related to the fact that Hawking radiation is semiclassically in a mixed (thermal) state and, due to the ``no hair'' theorems, carries no information at all about the collapsing body. As a consequence of this feature, a second loss regards the quanta of the field giving rise to the radiation. In Hawking's description, the outgoing modes of the field are correlated to the ingoing ones, even though one has no access to the latter: this is why the radiation state is thermal and it has an entanglement entropy associated to it. While this is a common picture in statistical mechanics---if you drop a box containing a gas in a trash can you have no access any more to the degrees of freedom inside the box, but the final state is still pure, as long as you take into account the whole system---, the problem with a black hole arises when this evaporates completely: the quanta in the radiation outside the hole are left in a state that is mixed, even though there is nothing to be mixed with anymore. In other words, the initial pure state has evolved into a final state that is mixed in a fundamental way.

Difficulties with the viability of Hawking's drastic proposal \cite{Hawking2} of a modification of the fundamental laws of quantum mechanics, in order to deal with this violation of unitary evolution, have been emphasized in \cite{Banks} (see, however, also \cite{Page2, Unruh}).

Another path often advocated as a possible way out of the paradox is to take into account some sort of non-locality \cite{Giddings}. In particular, this seems an unavoidable choice in string theory, where Bekenstein-Hawking entropy is interpreted in the strong form, as a measure of the number of degrees of freedom inside a black hole \cite{Vafa}\footnote{For a different perspective see \cite{Susskind2, Banks2}.}. The idea that non-locality could solve the paradox in string theory led to the introduction of notions such as ``black hole complementarity'' \cite{tHooft, Stret, Polchinski} and ``holographic principle'' \cite{Susskind}, which eventually culminated in the celebrated AdS/CFT correspondence \cite{AdS}. However, there are difficulties in finding evidence for such non-locality effects in string theory \cite{Giddings2} and no detailed analysis of the evaporation process, in either the CFT or the gravity theory, has explicitly been worked out, showing how locality breakdown can be reconciled with ordinary quantum field theory on a macroscopic scale. 

More recently, another picture within the string theory framework of how information can come out of black holes has been developed by modeling the hole in terms of so-called `fuzzball' states \cite{Mathur}. In such a picture, a traditional horizon never forms; instead, different states of the string (creating different fuzzballs) spread over a horizon sized region. In this way, the matter making the hole is not confined at the singularity, but fills up the entire horizon interior and the radiation emerging from the fuzzball can send its information out \cite{Mathur2}. However, also in this scenario reconciliation with local QFT in low curvature space-time regions is in a conjectural state. Moreover, extension of the fuzzball conjecture to the non-extremal case is a difficult task. 

In the LQG approach, the degrees of freedom responsible for the black hole entropy are located on the horizon (i.e., horizon quantum geometry fluctuations \cite{Rovelli}). Therefore, instead of locality, 
the assumption in Hawking's argument which is abandoned is the existence of an information-free horizon. 

In this section we are going to exploit this quantum hair structure associate to the fundamental discrete quantum horizon geometry to describe subtle modifications of the Hawking radiation. Moreover, by implementing the evaporation dynamics described in the previous section till the Planck regime is reached, we show how the classical singularity is resolved leading to the formation of a long-lived remnant at the end of the evaporation process. This represents a prediction following from the imposition of the LQG dynamics and is the main new result of this section.
We then explain how the combination of these two elements allows us to outline a picture for a possible solution to the information paradox in LQG. Furthermore, the role of non-local effects in the deep Planck regime due to non-local spin network links is considered, allowing for an extension of space-time beyond the classical singularity. 

In section \ref{Q-hair} we analyze more in detail the quantum hair notion emerging from the kinematical structure of the boundary Hilbert space. This section doesn't contain original material, but investigate the analogy between the LQG description of the IH quantum gravitational d.o.f. and the discrete gauge symmetries of black hole horizon studied in previous literature. Such analogy supports the microscopic understanding of black hole thermodynamics emerging from our analysis. In section \ref{Singularity} the massive remnant formation at the end of the evaporation process is derived using the dynamical evolution of IH quantum configurations described in section \ref{diss}. The robustness of this scenario and its implications for the information paradox are discussed in section \ref{Rem}, where possible implementation of non-local effects is also considered; this last section is of a more speculative nature.

\subsection{Quantum Hair}\la{Q-hair}

In the picture emerging from the LQG description of quantum black hole, the information resident on the horizon is encoded in a {\it quantum hair} at each puncture piercing the horizon. This notion of quantum hair, while of a very different nature, carries some similarities with the black hole quantum numbers associated with discrete $\Z_N$ gauge charge analyzed in \cite{Quantum-hair}. While compatible with (classical) ``no-hair'' theorems, the quantum hair considered by these authors, which are not associated with massless gauge fields, have semi-classical effects on the local observables outside the horizon and on black hole thermodynamics, affecting for instance the hole Hawking temperature. In particular, \cite{Quantum-hair} showed how, given two black holes with the same mass, the one with larger $\Z_N$ charge is cooler. As a consequence, quantum hair can inhibit the emission of Hawking radiation and therefore stop the evaporation process. We will show in the next section how the presence of a quantum hair associated to the quantum gravitational d.o.f. allows for a precise realization of such a stabilization mechanism by implementing the near-horizon quantum dynamics all the way till the hole reaches a Planck scale size. Moreover, the kind of quantum hair of the black hole analyzed in \cite{Quantum-hair} is argued to lead to non--perturbative corrections to the area law in \cite{Moss}, providing a further analogy with the LQG case.

Before analyzing the last stage of the evaporation process, let us make more explicit the parallel between the $\Z_N$ quantum hair considered in \cite{Quantum-hair} and the one introduced in the LQG framework. In \cite{Quantum-hair} the discrete $\Z_N$ gauge symmetry arises in the Higgs phase of a $U(1)$ gauge theory when a scalar with charge $Ne$ condenses. The residual $\Z_N$ subgroup which survives is related to the fact that the condensate cannot screen the electric field of a charge modulo $N$. Since the Higgs phase with unbroken $\Z_N$ local symmetry supports a `cosmic string', a vortex with magnetic flux $2\pi/Ne$ trapped in its core, the charge modulo $N$ on the black hole can be detected by means of the Aharonov-Bohm phase $\exp{(\i 2 \pi Q/Ne)}$ generated when a charge $Q$ is transported around the string. In this way, the $\Z_N$ electric hair induces an infinite range interaction between string and charge which has non-perturbative (in $\hbar$) effects on the dynamical properties of the hole. 
 
In LQG, as we recalled above, the quantum geometry d.o.f. on the horizon are described by a topological gauge theory with local defects \cite{ABCK, SU(2), SU(2)2}, namely by a Chern-Simons theory on a punctured two-sphere. As a result of the quantum implementation of \eqref{eom}, the punctures coming from the bulk and piercing the boundary represent the quantum excitations of the gravitational field on the horizon, as described by the LQG kinematics. If one adopts the point of view of \cite{Sahlmann}, the quantum version of (\ref{eom}) can be taken as the starting point for a full definition of quantum horizon within the LQG framework. In fact, the analysis of \cite{Sahlmann} provides a rigorous mathematical basis to realize the original intuition of \cite{Smolin} relating horizons in LQG to TQFT. The emerging picture is that of a quantum horizon as a {\it brane} of a flat connection with local excitations of the electric quantum field $\hat \Sigma$. Let us now see how the analog of the discrete gauge symmetry of \cite{Quantum-hair} arises in this context, when one restricts to the U(1) model. In \cite{Sahlmann} it is shown that the horizon quantum state $\Psi$ constructed as a solution of the quantum version of (\ref{eom}) is invariant under diffeomorphisms that keep the punctures fixed. However, this (gauge\footnote{In a topological field theory gauge invariance is strictly related to diffeomorphisms invariance.}) symmetry is broken if one considers the exchange of two punctures by diffeomorphisms that leave the other punctures invariant. This is a fundamental characteristic of the horizon state, since the distinguishibility of the punctures is a crucial property to recover the linear area behavior of the entropy, and it's a typical example of how the presence of a boundary can break the local symmetry and turn gauge d.o.f. into physical ones \cite{Carlip}. Nevertheless, there is a residual $\Z_k$ symmetry left at the punctures related to the fact that, by adding to the integers $m_p$ (labeling the $U(1)$ irreducible representations) associated to each puncture a multiple of the Chern-Simons level $k$, the horizon quantum state will not change. Such a symmetry, which is a well known property of Chern-Simons theory with punctures, derives from the boundary condition (\ref{eom}). In fact, by means of the Stokes theorem one has
\be\la{eom}
h_{\gamma} [A]= P \exp{\oiint}_{\!\!\!\!S}  F[A] = P \exp{\oiint}_{\!\!\!\!S} \frac{4\pi}{k}\Sigma,
\ee
where $h_{\gamma}$ is the holonomy around $\gamma=\partial S$; in the quantum theory then, when $\gamma$ goes around a puncture $p$ 
\be\la{holo}
\hat h_{\gamma}\Psi= e^{- \frac{2 \pi\i}{k}m_p}\Psi,
\ee
from which the local $\Z_k$ symmetry of the horizon spin network quantum state $\Psi$ follows. Notice that when the horizon has the topology of a 2-sphere, i.e. in the single intertwiner model, the punctures have to satisfy also the global constraint 
\be\la{inter}
\sum_p m_p=0~~{\rm mod}~~k,
\ee 
as a consequence of (\ref{holo}) when $\gamma$ goes around all the punctures on the horizon. 

It is now clear the analogy with the quantum hair considered in \cite{Quantum-hair}. Each puncture $p$ carries a $\Z_k$ electric charge given by its representation integer $m_p$; the Wilson loop for parallel transport around a puncture (\ref{holo}) defines an element of the residual gauge group $\Z_k$, measuring the flux inside with basic unit $\Phi_k=2\pi /k$. Therefore, the detection of the quantum hair effects by such observables is the analog of the Aharonov-Bohm interaction.

In \cite{Sahlmann} a Lebesgue measure for the integral on connections satisfying (\ref{holo}) is defined by fixing the $U(1)$ angle integration variable 
\be\la{phi} 
\phi_p=\frac{2\pi m_p}{k}
\ee
at each puncture $p$. In this way, a definition of `quantum isolated horizon' (QIH) within the full theory can be introduced. In fact, exploiting the analogy with the notion of quantum hair investigated in \cite{Quantum-hair}, the relation (\ref{phi}) can be interpreted as containing no reference to classical horizon elements: the integer $k$ can be assumed as a parameter proper of the topological quantum theory associated to a local, discrete quantum gauge symmetry; this definition of QIH would push further the point of view \cite{Sahlmann} and correspond to a realization at the  quantum level of the paradigm-shift introduced in \cite{Static, Static2}. Moreover, the extension of such a definition to the $SU(2)$ case, which requires a more involved analysis due to the highly nontrivial action of the operator on the r.h.s. of (\ref{eom}), seems to match the model of \cite{SU(2)} and  be compatible with the proposals of \cite{KR, LT}.

\subsection{Singularity Resolution}\la{Singularity}

We now want to exploit the existence of this horizon quantum hair and the associated description of the radiation emission of section \ref{diss} to explore the dynamics of the final stage of the evaporation process and its implication for the information loss. In particular, we want to investigate the idea that quantum geometry fluctuations may allow unitary evolution of the black hole beyond the classical singularity. The relevance of such a scenario for the resolution of the information paradox has been emphasized, for instance, in \cite{Ash-Bojo2, Hosse}.

In section \ref{diss} we saw that the quantization of the IH boundary condition \eqref{eom} yields a Chern-Simons theory on the 2-sphere horizon with punctures. In the large area limit then the boundary Hilbert space is isomorphic to a $SU(2)$ intertwiner space \cite{ABCK, SU(2), SU(2)2}. As we saw above, the approach of \cite{Sahlmann}, where the IH d.o.f. are represented by elements of the holonomy-flux algebra of LQG via a modified Ashtekar-Lewandowski measure on the horizon implementing \eqref{eom}, also converge to the single intertwiner model and shares the same topological features for the quantum boundary theory. Moreover, this picture is also compatible with the approaches \cite{KR, LT} attempting to provide a definition of quantum horizon without referring directly to the IH framework.

Therefore, let us concentrate on the $SU(2)$ intertwiner model for the IH quantum state and in particular on the more general approach of \cite{Sahlmann}. This can be constructed starting from an arbitrary spin network state on the canonical hypersurface $\Sigma$ with support on a graph $\Gamma$. We denote $B$ a connected bounded region of $\Gamma$ formed by a finite set of vertices and edges that connect them. The horizon surface $IH$ is defined as the set of $n$ edges with only one vertex in $B$,  which defines the boundary $\partial B$ of the connected region and is endowed with boundary data represented by the spin label $j_1,\dots, j_n$ of the boundary edges. To each puncture $p$ we can associate an elementary patch with a microscopic area given by the spin $j_p$: the collection of all these patches forms the quantum horizon brane. These colorings also determine the expectation values of observables defined on the horizon, in particular Wilson loops around the punctures are fully constrained by imposition of the boundary condition \eqref{eom} via the modified Ashtekar-Lewandowski measure on $IH$, as mentioned before. 
Moreover, in accordance with the expectation of the topological nature of the boundary theory, holonomies on all contractible loops in $IH$ do not represent local gauge invariant d.o.f. and the range of the spin labels is restricted by the Chern-Simons level playing the role of a cut-off in the quantum theory, namely $j_p\leq k/2$ \footnote{This expectation is also supported by the results of \cite{2+1} showing the emergence of quantum groups structures in the 2+1 theory.}. 

\begin{figure}[ht]
\centering
\includegraphics[scale=0.5]{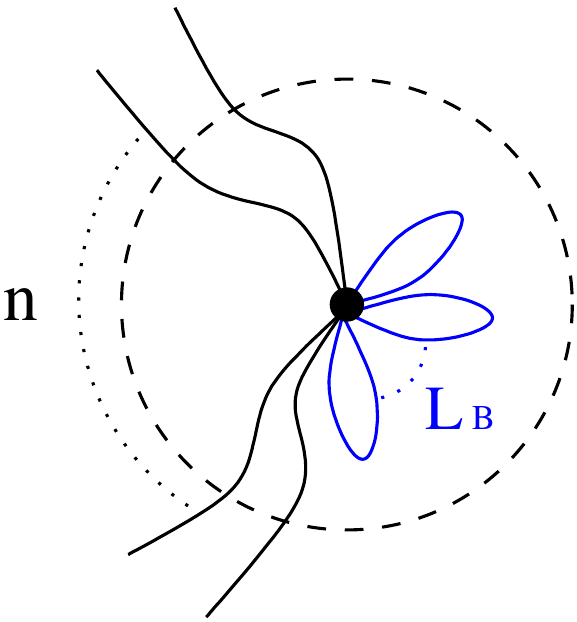}
\caption{General coarse-grained graph of a black hole interior with non-trivial topology.}
\label{Petals}
\end{figure}

The intertwiner structure can now be obtained in the following way.
The black hole interior geometry is fully described by the (superposition of) spin network states with support on the arbitrary complicated interior graph $\Gamma_B$. However, as argued above, for the entropy calculation the relevant information is only the one that can affect the external observer and be read off the horizon. Therefore, the details of the spin network inside the black hole do not matter and to an outside observer the horizon state looks like an intertwiner between the $SU(2)$ representations $V^{j_p}$. In other words, in the entropy calculation one traces over the bulk d.o.f. by coarse graining the interior geometry to a spin network having support on a graph with a single vertex inside the horizon, with the boundary edges coming out of it. The coarse graining techniques have been developed in detail in \cite{LT, LT2}, where it has been shown that the coarse-grained intertwiner in the boundary Hilbert space depends in general on the the coarse graining data $\{g_{e\in B/T}\}$, where $T$ is a maximal tree in $B$\footnote{A tree $T$ of $B$ is a set of edges in $B$ that go through all vertices of $B$ without ever making any loop.}. The group elements $\{g_{e\in B/T}\}$ represent the holonomies around each (non-contractible) loop of the interior graph $\Gamma_B$, once all the group elements on the edges belonging to the tree have been fixed to the identity, by means of gauge invariance. Henceforth, the coarse graining data encode the information about the non-trivial topology of the quantum geometry of the black hole interior $B$. Following \cite{LT}, having set the maximal number of holonomies to the identity, the whole graph $B$ can be effectively reduced to a single vertex with $n$ open edges (forming the boundary) and $L_B$ loops (see FIG. \ref{Petals}). The number of non-trivial loops of $\Gamma_B$ is given by $L_B=E_B-V_B-1$, where $|T|=V_B-1$ is the number of edges of the maximal tree $T$ in $B$ and $E_B$ the total number of edges in $B$. The quantum state of geometry of $B$ is then represented by the contraction of the single intertwnier $\mathcal I: V^{\otimes n}\otimes V^{\otimes 2L_B}\rightarrow \C$ with the holonomies $g_{e\in B/T}$ along the loops. Notice that the analog of the global constraint \eqref{inter} in the $SU(2)$ case corresponds to the condition that the sum of all the spin labels $j_p$ be an integer.

Let us now analyze how the evaporation process evolves this state by implementing the near-horizon dynamics as described in \cite{H-radiation}. We showed in section \ref{diss} how Thiemann's Hamiltonian operator acting on a node right outside the horizon with two edges departing from it and piercing the boundary surface creates a new link of spin $1/2$ (assuming that we regularize $\hat H$ in the fundamental representation) between the punctures and changes the spin associated to them. A typical process is when $\hat H$ acts on two spin-1/2 punctures and then one jumps to spin-0 while the other to spin-1 or on a spin-1/2 and a spin-1 punctures and they jump respectively to spin-0 and spin-1/2; graphically we have
\ba
&&\hat H \rhd \begin{array}{c}\includegraphics[scale=0.38]{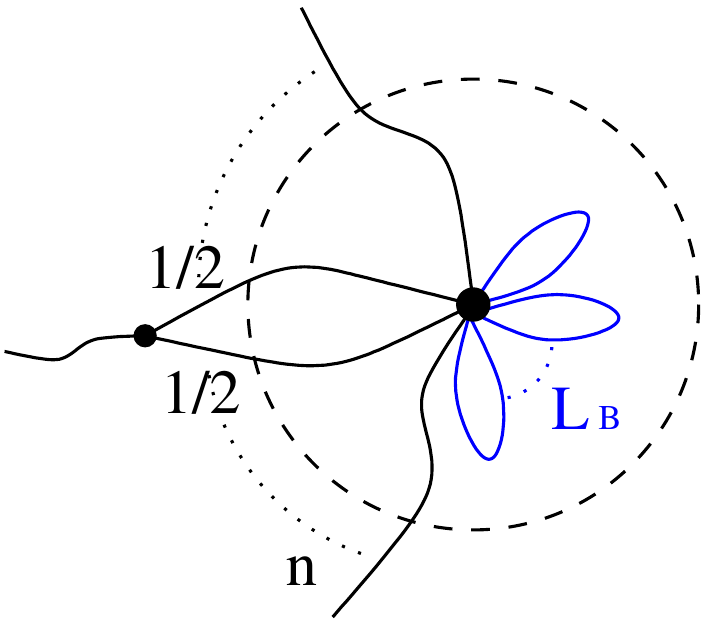}
\end{array}
~~\Rightarrow~~\begin{array}{c}\includegraphics[scale=0.38]{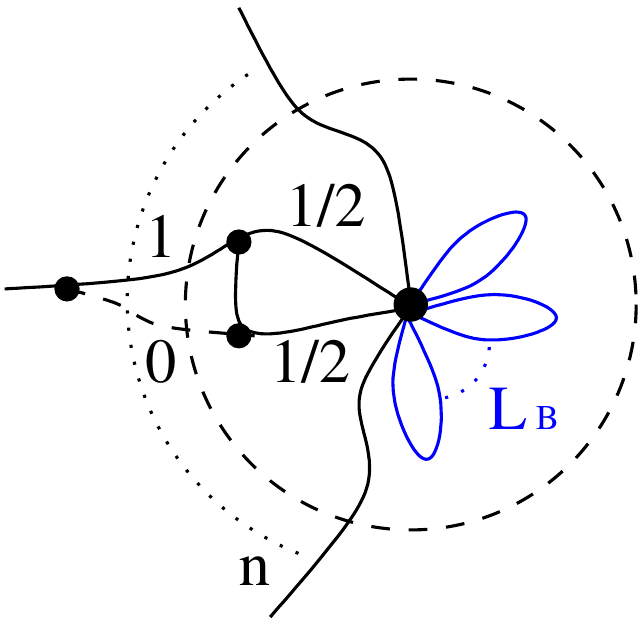}\end{array}=\begin{array}{c}\includegraphics[scale=0.38]{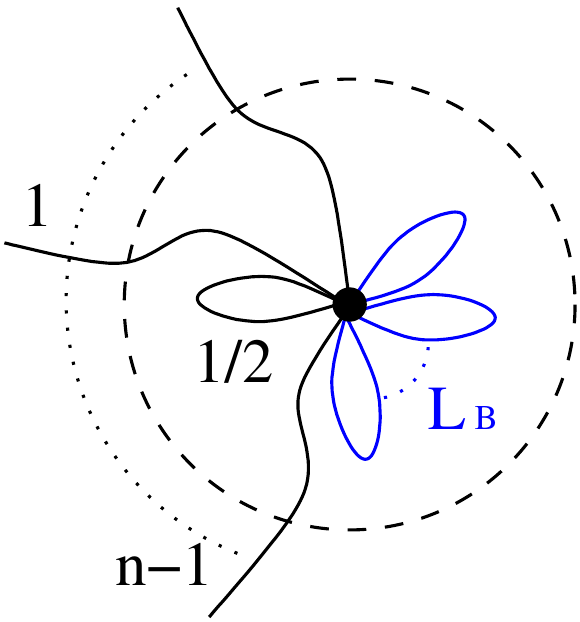}\la{petals1}
\end{array}\\
&&\hat H \rhd \begin{array}{c}\includegraphics[scale=0.38]{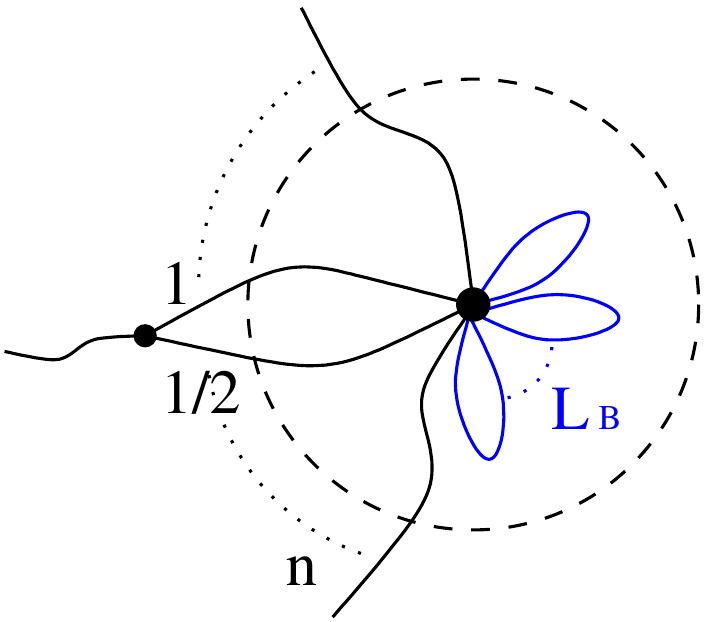}
\end{array}
~~\Rightarrow~~\begin{array}{c}\includegraphics[scale=0.38]{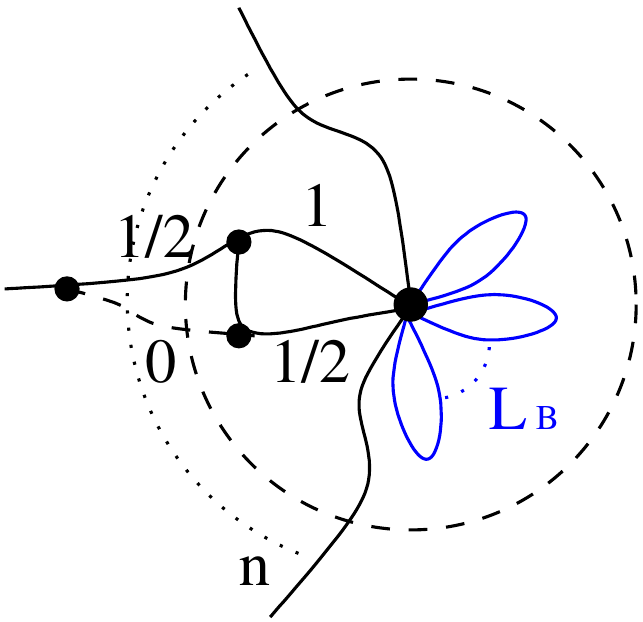}\end{array}=\begin{array}{c}\includegraphics[scale=0.38]{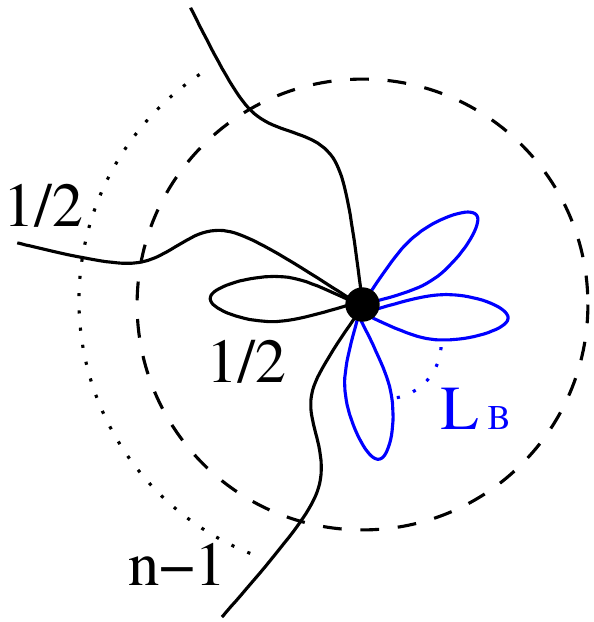}\,,\la{petals2}
\end{array}
\ea
where gauge invariance at the new trivalent node created inside the horizon has been used to form the petal. In this way, a boundary puncture disappears (the horizon shrinks) and a new loop of spin-1/2 is created in the interior of the hole every time. Of course, there will be also transitions in which the quantum geometry fluctuations correspond to an absorption process leading to an area increase. This for instance can happen when acting on a spin-1/2 and a spin-1 like in \eqref{petals2} and both punctures jump to an higher spin, namely 1 and 3/2. However, as pointed out in section \ref{diss}, assuming the standard area operator spectrum in LQG, the average process has a dissipative effect; this follows from the property of a decreasing gap between higher eigenvalues which makes most of the transitions induced by $\hat H$ correspond to an emission process. For example, among the six possible fluctuations allowed by the action of the Hamiltonian constraint for the initial configurations in \eqref{petals1} and \eqref{petals2} (which for a large black hole represent the most likely ones) only one is associated to an  absorption process, namely the one mentioned above. Obviously, the velocity of the evaporation process depends on the specific form of the Hamiltonian operator matrix elements and in general it is expected to be very slow for large black holes, consistently with the semi-classical picture.

We can now iterate this dynamics till the horizon reaches a Planck scale size in order to investigate the last stage of the evaporation process. Notice that in \cite{H-radiation} terms arising from the action of $\hat H$ where new punctures are created (corresponding to an absorption process with area increase) were discarded due to the breaking of diffeomorphisms symmetry on the horizon in those cases (for instance, such an action would create pathological states like those analyzed in \cite{Krasnov2}, making the horizon area observable ill defined). 
However, as the deep quantum regime is reached and the curvature becomes large, the notion of classical manifold breaks down.
 Therefore, in its final stage the horizon is expected to be in a quantum fluctuating state with no significant {\it quantum gravitational radiation} emission anymore:
boundary punctures are continuously being created and annihilated.
Such a final interior state carries resemblance with the picture of long hornlike geometries connected to the external space by tiny holes \cite{Banks2}, which classically would evolve to reach infinite length and zero width in finite proper time.

We can nevertheless wonder if the dynamics allows, at least in principle, a complete evaporation of the horizon where the area shrinks to zero, regardless of the velocity with which such a complete evaporation would take place. This can be investigated by implementing the action depicted in \eqref{petals1}, \eqref{petals2} as far as possible. Namely, assuming an initial distribution of only spin-1/2 and spin-1 punctures forming the boundary state, through a sequence of processes depicted in \eqref{petals1} and \eqref{petals2}, the quantum horizon could eventually reach a Planck scale state formed by only two punctures of spin-1/2 (notice that a Planckian state corresponding to two punctures of spin-1/2 and spin-1 is not allowed by the global gauge constraint mentioned above); if we now try to implement the action of the Hamiltonian operator \eqref{petals1} further in order to eliminate these residual two punctures, we realize that this is not allowed. The reason for this impossibility relies on the form of the matrix elements of (the Euclidean part of) $\hat H$ derived in \cite{Matrix}, where it is shown that the probability for two edges with the same spin $j$ to both jump to $j-1/2$ is zero. Therefore, only one of the two punctures can disappear inside the hole while the other has to jump to the spin-1 level, graphically
\be\la{final}
\hat H \rhd \begin{array}{c}\includegraphics[scale=0.38]{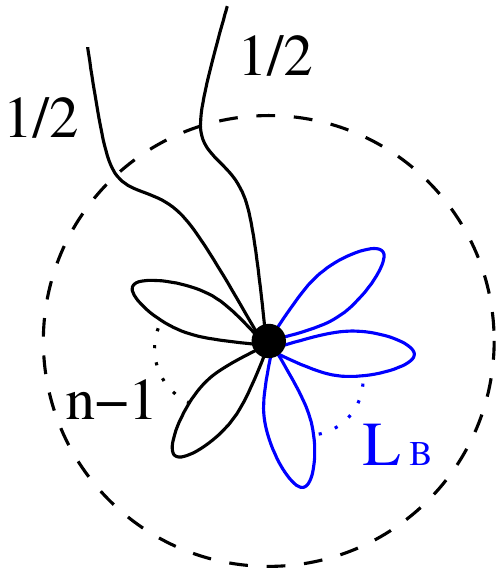}
\end{array}
~~\Rightarrow~~\begin{array}{c}\includegraphics[scale=0.38]{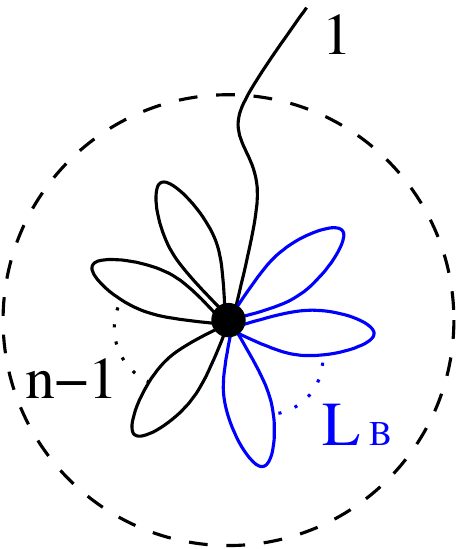}
\end{array}.
\ee
The (daisy-)state represented on the r.h.s. of (\ref{final}) corresponds to the allowed configuration of the quantum black hole with minimal area, showing how the horizon area operator is bounded from below by the dynamics of the theory, with a minimum value corresponding to $8\pi \beta \ell_p^2 \sqrt{2}$. Such an asymptotic state corresponds to a vanishing temperature limit of the evaporation process. A similar departure from the semi-classical scenario is found, for instance, also in \cite{Temp}, for the behavior of the surface gravity, and in \cite{Dark}, by means of dynamical arguments based on the generalized uncertainty principle.

Therefore, the analysis shows that the quantum horizon can never shrink completely, i.e. it never hits the ``$r=0$'' point (see FIG. \ref{Remnant} below). In this way, the spacelike singularity inside the black hole is removed due to the quantum dynamics of the theory, confirming the results of previous analyses \cite{Modesto, Husain, Singh, Ash-Bojo, Gambini} based on the application of mini-superspaces to Schwarzschild interior.

\begin{figure}[h]
\centerline{\(
\begin{array}{c}
\includegraphics[height=5cm]{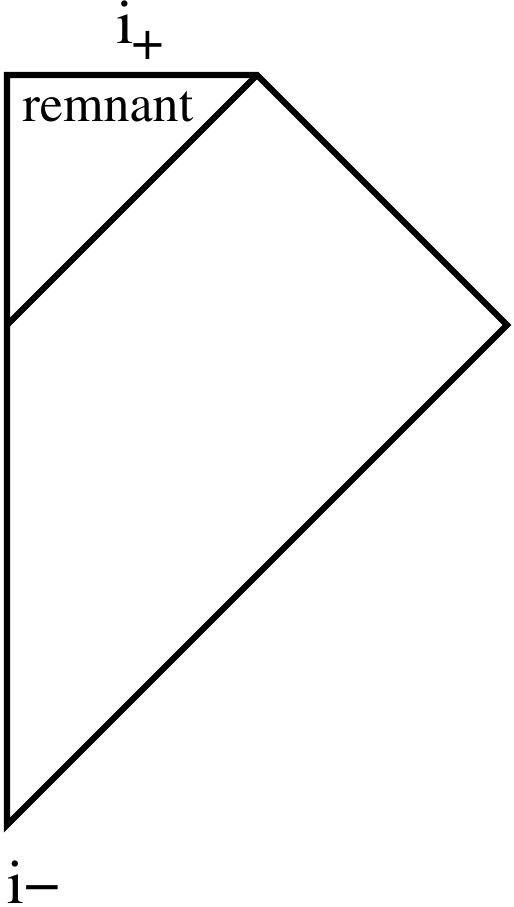}
\end{array}\!\!\!\!\!\!\!\!\!\!\!\! \!\!\!\!\!\!\!\!\!\!\!\!\!\! \begin{array}{c}  ^{} \, \,\, \,\, \,\, \vspace{1.8cm}      \sI^{\va +}_{}\\ \!\!\!\!\vspace{1.3cm} \sI^{\va -}\end{array}~~~~~~~~~~~~~~~~~~\begin{array}{c}
\includegraphics[height=3.5cm]{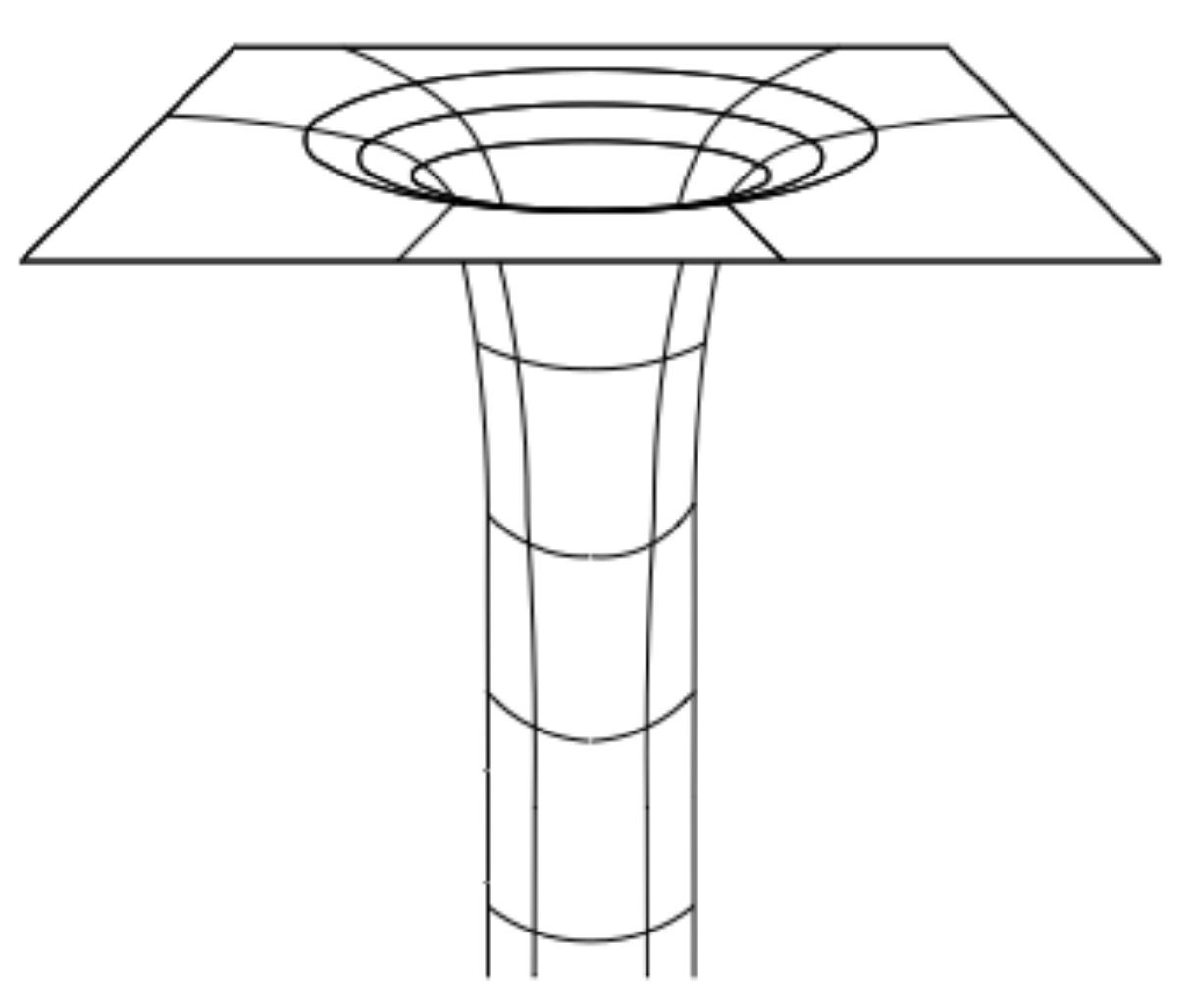}
\end{array}\) } \caption{On the left, the Penrose diagram of a black hole remnant. On the right, an angular slice of its static hornlike geometry. The horizon never shrinks down to zero size but stabilizes at a finite radius, forming a permanent massive remnant which never disappears from the original space-time.}
\label{Remnant}
\end{figure}

Notice that it is the `6--$j$' part of the euclidean Hamiltonian operator which is responsible of the avoidance of the complete evaporation of the horizon, as shown in (\ref{final}). Henceforth, this feature of the quantum gravitational description of the black hole evolution is not related just to Thiemann's regularization scheme and, in particular, it is  expected to be recovered also within the implementation the near-horizon dynamics in the spin foam formalism. In fact, since spin foam amplitude is expected to provide a definition of the physical scalar product in LQG, the fundamental action of $\hat H$ on a near-horizon node could be described by a vertex amplitude whose boundary graph pierces the horizon and contains the initial and final states as subgraphs. It is then easy to see how the part of the amplitude responsible for the disappearance of boundary punctures has the combinatorics of a 6--$j$. Moreover, the vanishing of the transition amplitude corresponding to the complete evaporation of the horizon holds for both orderings concerning the position of the volume operator inside the regularized expression for the Hamiltonian constraint, as considered in \cite{Matrix}; henceforth, the remnant formation scenario is free from this ambiguity.


We want to conclude this section with a remark concerning the first law. The identification in \cite{AP, APE} of the horizon area with a notion of local energy suggests a scheme for a microscopic derivation of it, within the framework we have been delineating. In fact, this local thermodynamics perspective is telling us that every flux of matter through the horizon leaves an imprint on it. From the microscopic theory, it is natural to see how this can be realized, since matter d.o.f. are associated to spin network links and, hence, matter falling into the black hole can happen only via the creation of new links piercing the horizon. This does not mean that the energy remains distributed on the horizon d.o.f.; indeed, it will fall inside and contribute to the interior state by creating new blu petals. However, 
the horizon Hilbert space will be modified by this flux and the surface system is left in a new ensemble. If now, according to the LQG recipe, we associate the black hole entropy to the dimension of the boundary Hilbert space\footnote{Notice that in this weak interpretation, it is actually {\em only} the imprint left on the boundary which contribute to the Bekenstein-Hawking formula.}, we see how the horizon can keep track of the entropy it gains as a bit of energy flows through and the first law follows naturally. 


\subsection{Black Hole Remnants}\la{Rem}

The last stage of the evaporation process described in the previous section shows how the horizon never shrinks down completely but stabilizes at a finite (microscopic) size. Such a final state corresponds to the formation of a massive remnant and leads to a non-singular quantum space-time according to the definition introduced in \cite{Hosse}, that is the dynamics defines a reversible linear map between the Hilbert spaces associated to two complete non-intersecting spacelike hypersurfaces. As argued in \cite{Hosse}, this property of the quantum dynamics is enough to restore unitarity once the d.o.f. inside the remnant are taken into account. 

\subsubsection{Where the information goes}

We can now see how information is not lost. First of all, let us point out how the d.o.f. associated to the collapsed matter that formed the black hole are encoded in both the boundary punctures forming the horizon and the coarse graining data associated to the non-trivial topology (the blue petals in FIG. \ref{Petals}) of the graph in the interior of the hole. The latter correspond to the large number of possible interior states, reflecting the variety of histories of the gravitational collapse, and, in the weak interpretation of the Bekenstein-Hawking entropy \cite{Sorkin, Smolin2, Jacobson}, they are not assumed to contribute since otherwise the result would be much larger.

The second form of information that is assumed to be lost in the semi-classical scenario consists of the correlations between the quanta of the matter field which reaches future null infinity via the Hawking process and their partners that fall inside the singularity. In our description of the evaporation process, the analog of these correlations are between the quanta of radiation that a stationary observer hovering close outside the horizon sees and the new petals that form inside the hole as the boundary punctures disappear in the emission process. 
However, the two types of information are strictly related since the boundary punctures data encode part of the collapsed matter degrees of freedom. 

A fundamental difference with the semi-classical description is that, for a local fiducial observer hovering at close distance from the horizon, the spectrum of the radiation is not thermal anymore, but formed by a discrete set of lines \cite{H-radiation}. While a smooth thermal envelope can be expected to be recovered for an asymptotic observer, for which the large number of punctures could compensate the suppression of more transition lines associated to small transition amplitudes, such a discrete structure would still emerge after a certain point from the beginning of the evaporation 
and before reaching the deep Planck regime. This is when the information associated to these correlations, and hence to the matter d.o.f., start to leak out.
More precisely, by measuring the energy levels and the intensity of the lines, by means of the spectrum formula (\ref{spectrum}), one can eventually recover the information about the microscopic structure of the quantum horizon encoded in the punctures data (i.e. how many punctures are in a given spin-$j$ level). Notice that the observation of such a spectrum would allow us also to solve ambiguities present in the quantization of the Hamiltonian constraint, like the irreducible representation to take the holonomies in and ordering ambiguities, as described in \cite{H-radiation}. However, to support this picture of information leakage a more detailed analysis of the radiation spectrum beyond the one-vertex approximation used in \cite{H-radiation} is required.

The rest of the information 
is stored behind the horizon of the fluctuating quantum final state of the evaporation process. No bit of information goes lost in the singularity, as the semi-classical analysis would suggest, since there is no singular final state anymore due to quantum dynamical effects. 

However, let us point out how the deviation from thermality associated to the quantum hair structure on the horizon could be enough also for an asymptotic observer to recover the full information at infinity. In fact, the quantum isolated horizon temperature has been recently derived in \cite{Temp} from a local microscopic analysis. The final formula presents a quantum correction associated to the Chern-Simons level, which defines an effective temperature reproducing exactly the deviation from thermality of the radiation spectrum found in \cite{Wil}. In \cite{Zhang} it has been argued how such a modification encodes correlations among quanta of Hawking radiation for a total amount of information corresponding to the Bekenstein-Hawking formula.

\subsubsection{Non-local effects}

A natural question then is whether this remnant state is permanent and the external observer will never have access to the information stored inside again (FIG. \ref{Remnant}) or it can actually decay, with the internal d.o.f. coupling again to the exterior spin network state, and the horizon disappear. 

 As we saw above, the blue petals in the interior of the black hole are associated to the non-trivial topology of the graph inside and encode part of the collapsed matter d.o.f. that formed the horizon (or keep falling inside after its formation) and do not leave a direct imprint on the surface state. Therefore, they are to be interpreted as bulk d.o.f. and, in its weak interpretation, they do not contribute to the Bekenstein-Hawking entropy. Nevertheless, they may play an important role in the black hole evolution, since, due to quantum fluctuations of geometry induced by the dynamics, it could be possible for them to tunnel out (see FIG. \ref{exit}) and, given that the boundary conditions defining the quantum horizon (no matter their specific choice) may not be preserved by this dynamical evolution violating its causal structure, to destroy the horizon. In this way, the space-time can extend beyond the classical singularity, corresponding to a sector of the phase space where the triad has reversed orientation. All the information that was trapped within the apparent horizon could eventually get out to infinity by means of the time-reversed action of $\hat H$ driving the evaporation process.
However, given the local nature of the Hamiltonian operator action and the very large volume of the bulk region (due to the presence of a big intertwiner), one would expect the amplitude for this tunnel effect to be considerably suppressed and such a scenario highly unlikely. 
 
\begin{figure}[h]
\centerline{
\(\begin{array}{c}
\hat H \rhd \end{array}
\begin{array}{c} \includegraphics[scale=0.37]{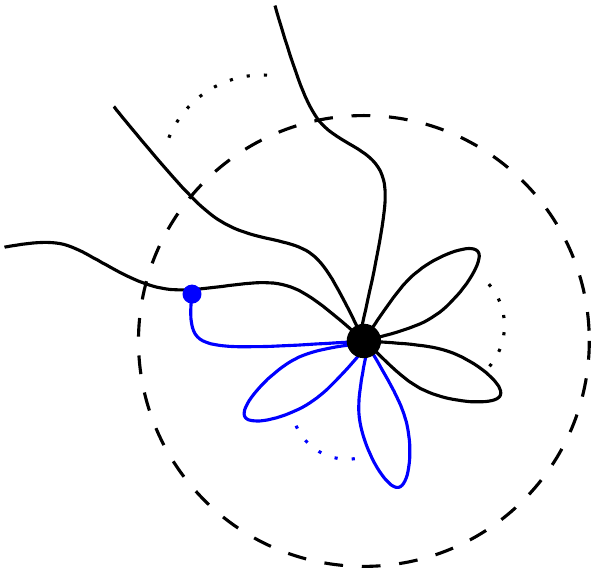}\end{array}
\begin{array}{c}~~\Rightarrow~~\end{array}
\begin{array}{c}\includegraphics[scale=0.37]{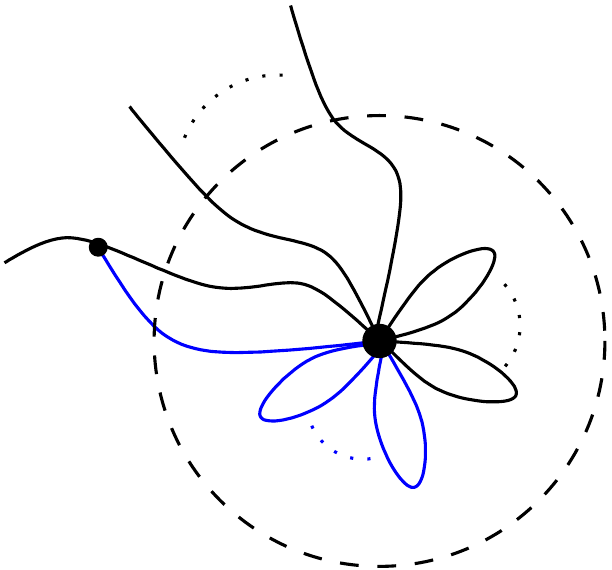}\end{array}\)}
\caption{Example of a dynamical process allowing the d.o.f. inside the horizon to tunnel out.}
\la{exit}
\end{figure}

\begin{figure}[h]
\centerline{\(
\begin{array}{c}
\includegraphics[height=6cm]{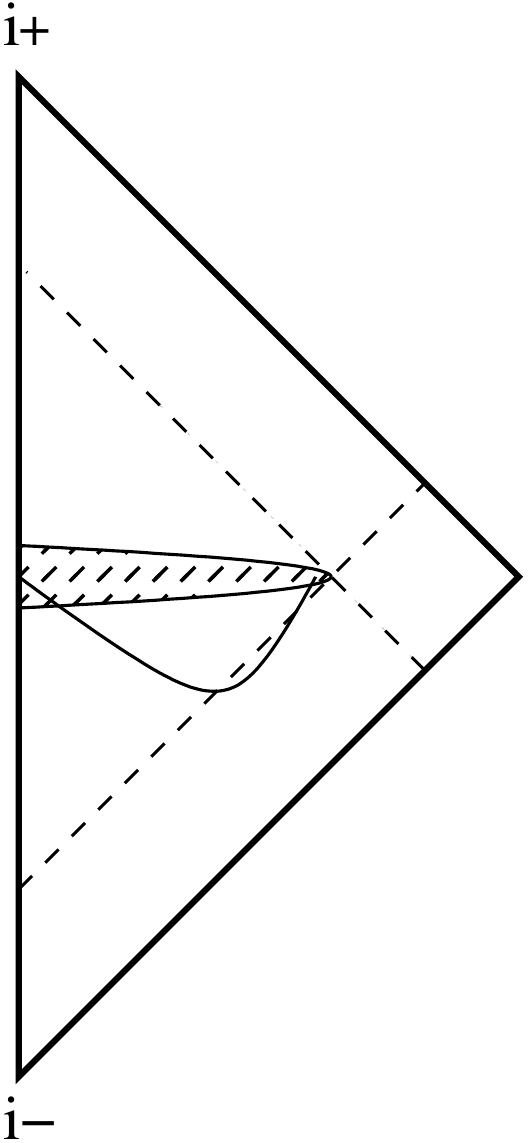}
\end{array}\!\!\!\!\!\!\!\!\!\!\!\! \!\!\!\!\!\!\!\!\!\!\!\!\!\! \begin{array}{c}  ^{} \, \,\, \,\, \vspace{2.7cm}   \hspace{-.6cm}   \sI^{\va +}_{}\\\!\! \!\!\!\!\!\!\vspace{.2cm} \sI^{\va -}\end{array}
\begin{array}{c}  ^{}\vspace{-1.1cm}   \hspace{-1.7cm}   \sH\end{array}
\) } \caption{Space-time diagram of an unstable remnant. The dynamical horizon $\sH$  will first grow during collapse, it will  eventually settle down to an isolated horizon and then slowly shrink till it reaches a minimum area value.
The shaded region corresponds to the deep Planck regime in which non-local effects might eventually lead to the horizon dissolution. In this way space-time extends beyond the classical singularity and the information trapped inside the remnant can now get out and reach future null infinity $\sI^{\va +}$.}
\label{Remnant3}
\end{figure}

At this point, an important observation concerns the notion of locality for the graphs underlying the 3-geometry quantum states. In \cite{Fotini} (see also references therein) it has been shown how the notion of {\it microlocality}, associated to the connectivity of the combinatorial structure of the graph, and that of {\it macrolocality}, associated to an emergent classical space-time metric, need not to coincide in the definition of semi-classical states\footnote{Notice that, due to the dominating contribution to the quantum horizon states of links with the lowest irreducible spin-representations, the combinatorial and geometric notions of locality near the horizon coincide.}. The authors show that states which are semiclassical but nonetheless contain non-local links are common in the physical Hilbert space of LQG. Moreover, it is argued that, starting with a local state associated with a classical three metric and implementing a long series of local moves induced by the LQG dynamics, one can create non-local links, which are not suppressed by further implementation of dynamics. This suggests a concrete realization of non-local effects in the evaporation process we described, which does not violate the local quantum field theory description of physics at low curvatures. Namely, the black hole bulk state can start out without any non-local links, allowing a local low energy limit of the quantum gravity theory in space-time regions where the semi-classical approximation is supposed to be valid. However, in the long term, over time scales large compared to the Planck time (as the horizon shrinks down and the curvature becomes big), non-local connections are introduced in the interior state. In this way, it would be possible for the in-falling initial d.o.f. in the deep Planck regime to tunnel out, as depicted in FIG. \ref{Remnant3}.

Henceforth, the presence of these non-local effects induced by the LQG dynamics, while compatible with the semi-classical analysis in its regime of validity, can
provide a concrete example of the notion of non-locality invoked in \cite{Giddings3}. This is surely a scenario that deserves further investigation and at this stage it has to be taken just as a proposal for a possible realization from the full theory of the paradigm described in \cite{Ash-Bojo2, Hayward}.

\subsubsection{Objections against remnants}

Problems with permanent or long lived remnants have been raised and discussed by several authors in the literature (see for instance \cite{Preskill, Remnants, Remnants2, Banks2, Page2}). Some of these objections haven been addressed also in \cite{Hosse}. Our description weakens further the criticisms to the remnants scenario, the main one being the infinite pair production problem. Let us address this issue more in detail.

First of all, the dynamical nature of the horizon and bulk states described above, due to quantum geometry fluctuations allowing for the internal d.o.f. to couple with the external ones, precludes the application of effective field theory with minimal coupling to remnants, as argued in \cite{Hosse, Banks2}. Treating black hole remnants as pointlike particles is not allowed, once quantum gravity dynamics is taken into account, and the tensor product structure of the Hilbert space becomes fuzzy in this regime. Another important observation that would invalidate an effective QFT treatment of the remnant is related to the modified first law \eqref{first}. In fact, as noted by L. Freidel, such a modification, necessary in order to take into account dissipative effects on the horizon (as discussed in Section \ref{Sciama}), could allow the remnant to have a mass well above the Planck mass.

Moreover, despite the fact that the formation of a big intertwiner in the interior of the remnant state allows for a region with very large volume (hidden behind the small horizon), the amount of information to be stored inside the remnant, in order to restore unitarity of the black hole evolution, is not as large as usually expected.
In fact, as argued above, the radiation process we described allows to recover part of the information about the initial collapsed matter d.o.f. and the radiation correlations already before the evaporation process stops. This means that the infinite degeneracy of the interior state advocated to affect the effective field theory description of the coupling of the remnants to soft quanta is not such a valid objection in our case. By the time the remnant forms, an external observer has already had access to part of its internal structure states and the amount of information left to recover in order for the Bekenstein-Hawking entropy (in its weak interpretation) to vanish is not as large to require an infinite degeneracy of remnant species.

\section{Conclusions}\la{Conclusions}

If we adopt the point of view that the entropy d.o.f. reside at the horizon and we refer to a local description of black hole thermodynamical properties, a theory of quantum gravity can be successfully applied to shed light on the fundamental problems of black hole physics. 
We have seen how the interpretation of a quantum horizon as a gas of punctures (`atoms' of space) whose microscopic dynamics is described by the LQG formalism, provides a thorough statistical mechanical analysis of the system. 

In \cite{Sciama} it was shown how Hawking radiation can be recovered entirely in terms of the viscous nature of the horizon associated to purely gravitational d.o.f.; if we take this indication seriously, at the local microscopical level then one should be able to describe the evaporation process by focusing exclusively on the dynamics of the quantum geometry constituents. 
In particular, the horizon evolution, when described in terms of a fluctuation-dissipation relation applied to the quantum hair associated to the fundamental discrete structure, provides a description of the radiation emission in terms of relaxation to an equilibrium state balanced by the excitation of Planck scale d.o.f. at the horizon. In this way, the space-time dissipative effects, encoded in a modification of the first law \eqref{first}, are related to the quantum geometry fluctuation induced by the Hamiltonian operator, providing a connection between macroscopic and microscopic levels of descriptions. This new description of the evaporation process is one of the two main results presented in this paper and it has important implications for the information paradox.

Namely, a fiducial observer hovering very close to the horizon and capable to perform measurements with sufficiently fine time resolution could reconcile the effective `viscosity' of the horizon with the unitarity of the process. In fact, such an observer would be able to access most of the information of the matter d.o.f. that fell inside the black hole and left an imprint on the horizon from the details of the radiation spectrum she observes on extremely short distance and time scales. An asymptotic observer, however, can only discuss average properties of the hole and have access to the information only after a long time from the beginning of the evaporation process. 

It is this broadening of the spectrum lines, which is expected to take place in the initial phase of the evaporation for large black holes, the coarse-graining procedure necessary to prove the second law. In this regime, the distant observer would not be able to read off the correlations between the emitted quanta and the in-falling d.o.f. from the radiation spectrum, with the entropy increasing in time due to this constant injection of entanglement. However, as the horizon shrinks down in size, the spectrum starts to reveal its discrete structure and the correlations eventually become detectable. This is the point when the black hole entropy curve flip over and start descending, in a scenario similar to the one envisaged by \cite{Page}, where an estimation of the information contained in the Hawking radiation subsystem (forming a random pure state with a second subsystem represented by the black hole) shows that information could indeed gradually come out via correlations between early and late radiation parts. 

In any case, the information that could not be recovered from the radiation content is not lost. In fact, 
when taking into account the local quantum dynamics of the gravitational field, black hole evolution is not singular and unitarity can be recovered. More precisely, 
a quantum mechanical description in terms of Hilbert space structures of the black hole evolution is valid and possible at all the stages. After the deep Planck regime is reached and part of the information has leaked out through the spectrum of the quantum gravitational radiation, the collapsed matter forms a massive remnant, which does not radiate anymore and whose d.o.f. are described by a separate Hilbert space. This is the second main result of our analysis. In this high curvature regime, if no non-local effects take place, the interior state information, whilst not lost, won't be able to come out and be accessible again to an exterior observer. On the other hand, if non-local effects of the kind described by the notion of {\it disordered locality} \cite{Fotini} develop, the horizon could dissolve in this quantum gravitational phase, with the consequent vanishing of the trapped surface (see FIG. \ref{Remnant3}). In this case, all the d.o.f. on a complete Cauchy surface can eventually be described again in terms of a single Hilbert space. Inclusion of this non-local effects and their possible connection with the complementarity ideas introduced in the string theory literature deserve and necessitate of a more detailed investigation.  

 While each of these solutions, proposed at different stages in the literature, faces serious problems when taken singularly, the combination of them, allowed and actually realized by the unitary dynamics of LQG, provides a valid alternative to more drastic departures from semi-classical physics. 

Surely, the analysis presented here is far from conclusive. Among other things, a more thorough investigation of the radiation spectrum derived in \cite{H-radiation} has to be carried out; a precise canonical definition of the quantum horizon from the full theory needs to be sort out, possibly allowing to derive a notion of Unruh temperature in terms of the horizon geometrical d.o.f.; matter should be included in the picture and the gravitational imprinting referred to at the end of Section \ref{Singularity} described in detail; the Lorentzian part of the Hamiltonian constraint should be added to the derivation of the radiation spectrum and its implications for the formation of a massive remnant analyzed in order to make this scenario more robust.
Nevertheless, we believe that the picture presented here provides a coherent and appealing description of the statistical mechanics understanding of black holes thermodynamics.

Let us conclude by pointing out that our analysis provides a concrete realization of the conjectured scenario \cite{Dark2} in which black
hole evaporation could cease once the hole gets close to
the Planck mass, allowing for the formation of Planck relics which could contribute to the dark
matter (see also \cite{Dark3} for a more recent proposal along these lines).
Furthermore, the insights gained in the context of black holes on the interplay between microscopic and macroscopic scales might turn out to be useful in application of emergent space-time scenarios to the investigation of the semi-classical continuum limit of the theory \cite{Oriti}.


\vskip0.3cm
{\it Acknowledgements.} I would like to thank E. Bianchi, B. Dittrich, D. Oriti, A. Perez, L. Sindoni and L. Smolin 
for useful discussions and comments which helped to improve this manuscript. Comments from an anonymous referee have also contributed to make the presentation clearer. 


\end{document}